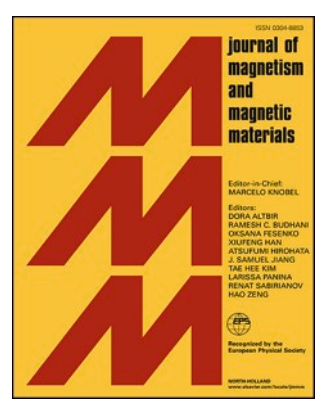

Research article

# Influence of chemical ordering on magnetocrystalline anisotropy and magnetoelastic properties in Weyl magnetic semimetal $Co_2MnGa$ thin films

O.M. Chumak [a], A. Nabiałek [a,*], T. Seki [b], J. Wang [b,c], K. Takanashi [b,d], L.T. Baczewski [a], H. Szymczak [a]

[a] *Institute of Physics PAS, Al. Lotników 32/46, 02-668 Warsaw, Poland*
[b] *Institute for Materials Research, Tohoku University, 980-8577, Miyagi, Japan*
[c] *Innovative Functional Materials Research Institute, National Institute of Advanced Industrial Science and Technology, 463-8560, Aichi, Japan*
[d] *Advanced Science Research Center, Japan Atomic Energy Agency, 319-1195, Ibaraki, Japan*

ABSTRACT

The magnetic anisotropy and magnetoelastic properties of the magnetic Weyl semimetal $Co_2MnGa$ were investigated in the $Co_2MnGa$ thin films with different chemical orderings. Their properties change, in line with the increase in the contents of the highly-ordered B2 and $L2_1$ phases. The most significant changes of both the magnetic anisotropy and the magnetoelastic properties were correlated with the appearance of the $L2_1$ ordered phase of Heusler alloy, which is also responsible for topological properties of these compounds. The magnetoelastic properties of $L2_1$ ordered $Co_2MnGa$ are discussed in comparison with the $Co_2Fe_xMn_{1-x}Si$ system with a similar degree of ordering, which has similar crystallographic structure but it is not Weyl semimetal. In particular, significant differences between the magnetoelastic properties of ordered $Co_2MnGa$ and $Co_2MnSi$ were found. In the case of $L2_1$-ordered $Co_2MnGa$, the giant anisotropy of the magnetoelastic properties was observed.

## 1. Introduction

Some full Heusler alloys of the $X_2YZ$ type (where X and Y are transition metals) have been theoretically predicted to exhibit ferromagnetic Weyl semimetal properties [1,2]. Among them, the Weyl semimetallic feature has been experimentally confirmed in bulk single crystals of $Co_2MnGa$ (CMG) through the detection of linear band dispersions using photoemission spectroscopy [3]. Epitaxial thin films of a CMG Heusler alloy with topological properties have gained significant attention due to their potential applications in spintronics and spin-caloritronics given that the CMG films exhibit the large anomalous Hall effect (AHE) and anomalous Nernst effect (ANE) [4,5,6]. In addition, given the potential use of AHE and ANE in future technologies, e.g., as a flexible device [6], understanding the magnetic anisotropy and magnetoelastic properties of these films is crucial for optimizing their performance in practical applications.

The Co-based full Heusler alloys exhibit different phases: the $L2_1$ phase with a perfectly ordered atomic arrangement, the B2 phase with Y–Z atom displacements, the $D0_3$ phase with Co—Y displacements, or the A2 phase where the Co, Y, and Z elements are randomly distributed [7]. According to the previous work [8,9], the degree of chemical ordering in the $L2_1$ structure of CMG is related to the emergence of magnetic Weyl semimetallic feature and also plays a key role in determining its magnetic properties [10,11], including the direction of easy magnetization and the magnitude of magnetocrystalline anisotropy (MCA). The electronic structure is varied by the chemical ordering, leading to changes in magnetic interactions and strain-induced effects. Studying these properties in thin films with different degrees of chemical ordering can provide insight into how structural factors influence magnetism at the atomic level. This is particularly relevant for improving the stability and tunability of magnetic anisotropy, which are essential for sensor technologies, memory devices, and spintronic applications.

In a recent study, it was found that the giant anisotropy of spin-orbit torque having its origin in the interplay between the topology of the electronic states and the strain exerted on the films by the substrate with non-identical lattice parameters in the $L2_1$ ordered CMG films [12]. This result shows that the tetragonal strain can modify the spin-orbital interactions even in the case of highly symmetric $L2_1$ CMG. For this reason, it seems even more interesting to study how the change of chemical ordering of CMG affects its magnetoelastic properties.

* Corresponding author.
*E-mail address:* nabia@ifpan.edu.pl (A. Nabiałek).

The previous work [13] reported the effect of disorder and vacancy defects on the electrical transport properties of CMG thin films. It was shown that by the application of different annealing temperatures, the long-range degree of chemical ordering was changed. Samples annealed at 300 °C contained only the A2 phase. Increasing the annealing temperature to 400 °C and further to 500 °C leads to the appearance of ordered phases B2 and $L2_1$, respectively. Also, a remarkable strain-induced MCA was reported for the epitaxial CMG films [14], and the magnitudes of both perpendicular to the plane and in-plane MCA increase with the layer distortion [15], which suggests the magnetoelastic effect to be significant.

Obtaining a perfectly ordered $L2_1$ phase under realistic technological conditions can be very difficult. Therefore, from a practical perspective, it is extremely important to understand how admixtures of lower-symmetry phases change the properties of such materials. This is particularly true for thin films, whose practical applications seem to be the most promising.

In practical applications of thin films, the ability to control and, if necessary, modify the magnetocrystalline anisotropy is extremely important. The change of anisotropy of thin layers can be achieved using the magnetoelastic effect and appropriate matching of the lattice constants of the magnetic layer and the substrate.

There is also a relationship between the magnetoelastic effect and damping of electromagnetic waves at microwave frequencies [16,17], which is crucial for applications in electronics.

For the reasons mentioned above, it is very important to know the magnetoelastic properties of thin films, and it is known that these properties may differ significantly from the properties of bulk materials of the same composition [18].

There are few methods for studying the magnetoelastic properties of thin films, and one of them is the strain modulated ferromagnetic resonance (SMFMR) [19,20], which is used in our present studies.

The correlation between the chemical ordering and the magnetoelastic properties in CMG has not been studied so far. Hence in this paper, we focus on the correlation between the chemical ordering, MCA and magnetoelastic properties in the epitaxial CMG thin films. Our previous results show that the magnetoelastic properties of the epitaxial $Co_2Fe_xMn_{1-x}Si$ Heusler alloy thin films can be strongly anisotropic [21] and the strain determined the magnitude of MCA [22]. Unlike CMG, these compounds are not Weyls semimetals. For this reason, comparison of the properties of these two groups of materials may shed some light on how the topological properties of CMG can affect the anisotropy and the magnetoelastic properties.

## 2. Experiment

The 75 nm thick epitaxial $Co_2MnGa$ layers were grown directly on MgO (001) and protected by 2 nm Al cap layers. The Al capping layer was next naturally oxidized in air and stable aluminum oxide layer was formed. This procedure allows to avoid a formation of the ternary compound $Co_2MnAl$ during the annealing process.

In the case of $Co_2MnGa$ films, the use of Al capping layers poses a risk of interfacial reaction between the Al cap layer and the $Co_2MnGa$ film and the possible formation of parasitic $Co_2MnAl$ compound. In contrast to pure Al, aluminum oxide is more stable chemically, which excludes the possibility of forming parasitic phases. In previous studies [13], our samples were examined by high resolution transmission electron microscopy (HR TEM), energy-dispersive x-ray (EDX) analysis, x-ray diffraction (XRD) analysis, magnetization, (M–H) measurements at room temperature, and Hall effect measurements at 10–296 K. No evidence of $Co_2MnAl$ parasitic phase was found. It is also worth to underline a very large difference in thicknesses between $Co_2MnGa$ film (75 nm) and Al capping layer (2 nm) what significantly reduces the possibility of a parasitic phase (if any) formation and its influence on the studied properties.

After oxidation of the Al capping layer in air, the samples were annealed at different temperatures ($T_a$) in the range from 300 °C to 500 °C in vacuum for 3 h. We present here the results for $Co_2MnGa$ samples annealed at 300 °C, 400 °C and 500 °C. The details of sample preparation were described in the previous paper [13], in which the structural properties were reported using TEM, EDX analysis, XRD analysis, as well as positron annihilation measurements.

In this paper, the measurement of anisotropic magnetoelastic properties of the CMG thin films was performed using the strain modulated ferromagnetic resonance (SMFMR) technique [19,20]. It is one of the methods that allows the study of anisotropic magnetoelastic properties of thin films, while the conventional ferromagnetic resonance (FMR) technique is a method to accurately determine the magnetic anisotropy. Hence, the conventional FMR technique was used to determine the magnetic anisotropy of the thin films.

In the FMR studies, we used the X-band Bruker EMX EPR spectrometer. The maximum field attainable in this system was about 18 kOe. This system was used to study the angular dependencies of the resonance field in the in-plane and out-of-plane directions, which were used to calculate the cubic and uniaxial MCA constants perpendicular to the film plane, respectively.

The SMFMR studies were performed using an X-band EPR spectrometer modified for this purpose, with a maximum attainable field of about 9.5 kOe. Hence, only the in-plane studies could be performed to determine the two magnetoelastic constants characteristic of the samples with cubic symmetry. The procedure to determine the magnetoelastic constants was the same as that described in our previous paper [21]. In the SMFMR technique, we compare the magnitudes of two FMR signals, one of which is modulated by an AC (100 kHz) magnetic field and the other one by periodic (48 kHz) strain [19,20,21]. The comparison of these two signals enables the determination of the shift of the resonance line caused by the strain, and then, using an appropriate model, the calculation of the magnetoelastic constants. Using this method, we can determine two magnetoelastic constants characteristic for the samples with cubic structure (constants $b_1$ and $b_2$ - see Eq. 3).

The magnetization of the films was measured at room temperature using the vibrating sample magnetometer (VSM) in the magnetic field range of $+/-$ 2 kOe. Using these data, we can determine the saturation magnetization of magnetic layers, which is necessary in our analysis.

## 3. Results and discussion

The structure of the samples studied here was already carefully characterized. According to the previous paper [13], after annealing at 300 °C only the A2 phase was detected. The annealing at 400 °C dramatically increased the amount of the B2 phase (up to 87%), and only after annealing at 500 °C the $L2_1$ phase (23%) in addition to the B2 phase appeared. Higher annealing temperatures did not significantly increase the content of $L2_1$ phase which is why we have studied the three samples with the most pronounced structural differences. The CMG films contained divacancies at a concentration of at least 100 ppm, and these were partially annealed out at a temperature above 400 °C near the interface with MgO [13]. The results reported in Ref. [13] also showed that after increasing the annealing temperature up to 500 °C both longitudinal and anomalous Hall conductivities were dramatically enhanced, although they remained lower than for the bulk materials.

A similar degree of ordering (about 80% of B2 and 20% of $L2_1$) was obtained for the $Co_2Fe_xMn_{1-x}Si$ layers we studied earlier with an iron content of $0 < x < 1$ [21]. Therefore, a comparison of these two systems may be particularly interesting. Because it gives an opportunity to determine the influence of other parameters, including topological properties, on magnetoelastic anisotropy and magnetoelastic properties.

The magnetization hysteresis loops of the three samples are shown in Fig. 1(a-c). They were measured in an external magnetic field with the range of $+/-$ 2 kOe, in which the magnetic field was applied along the in-plane [100] axis or [110] axis of the CMG layer.

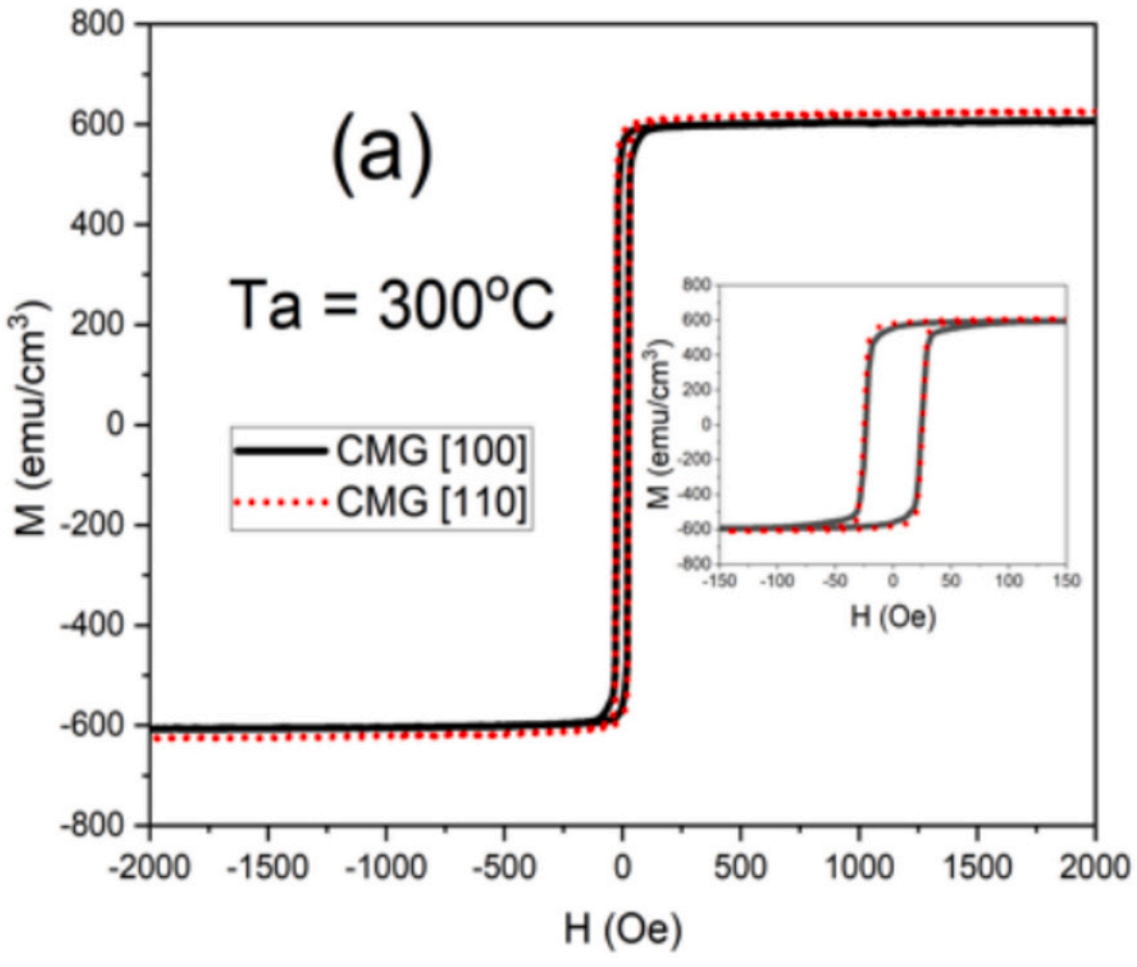

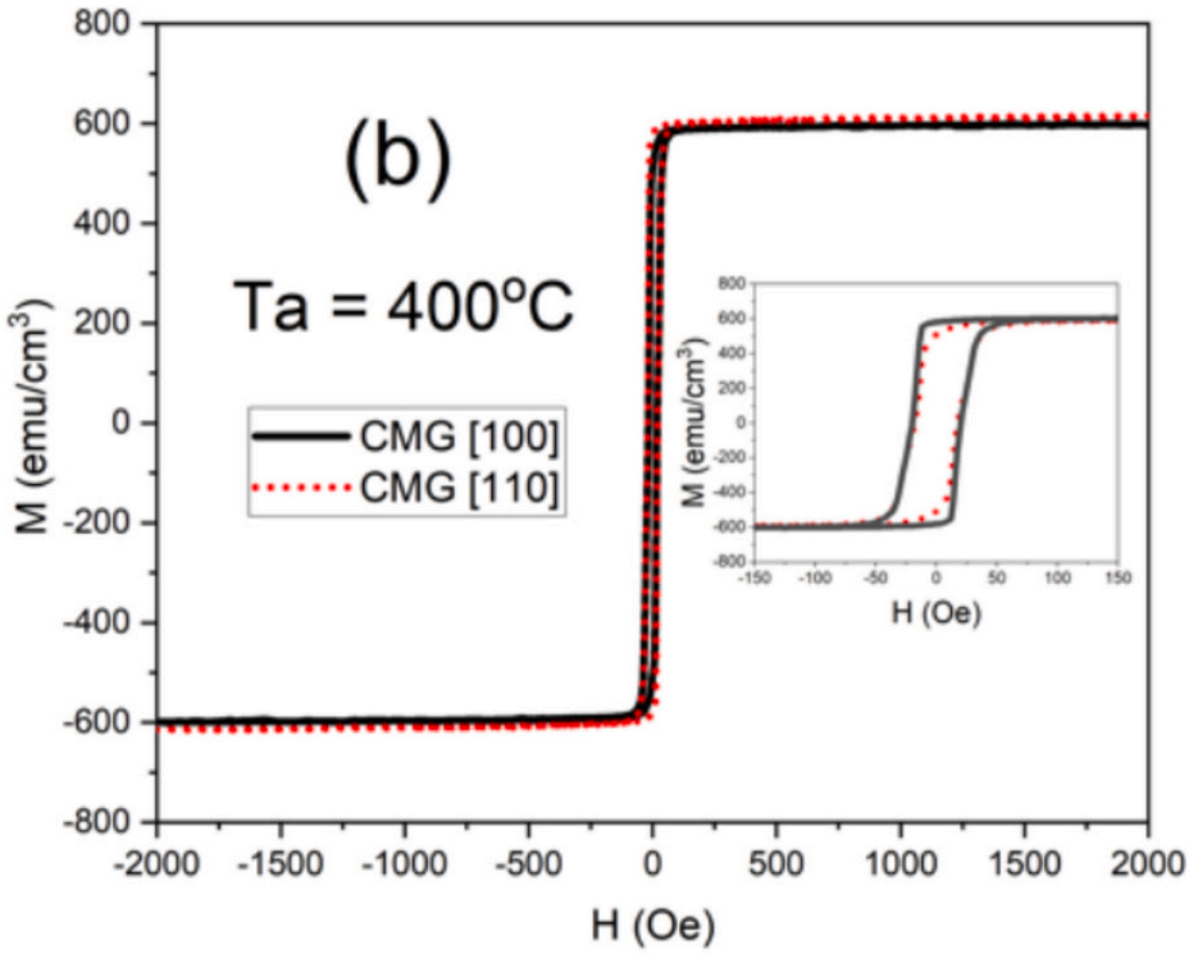

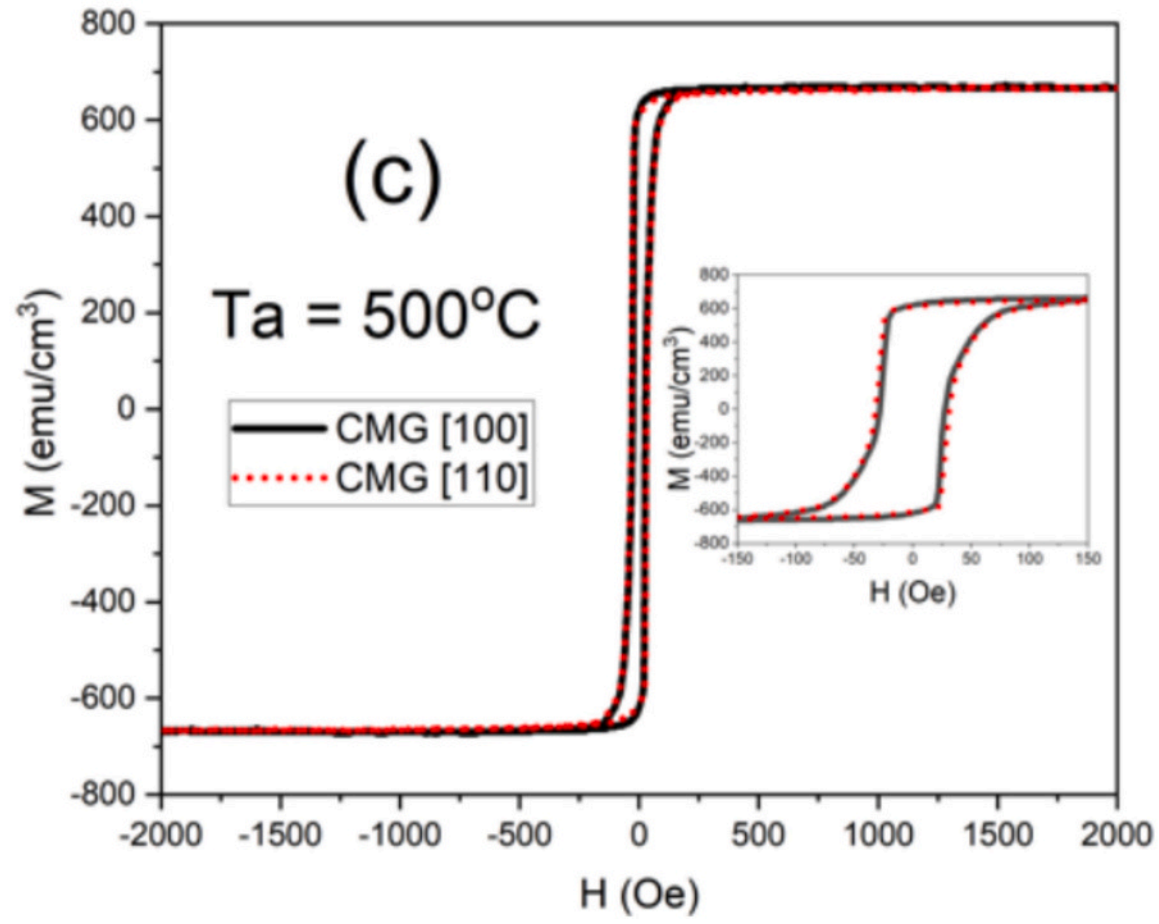

**Fig. 1.** Magnetization hysteresis loops for the three samples. The samples were annealed at $T_a$ Ta = 300 °C (a), 400 °C (b) and 500 °C (c). The loops were measured at room temperature, and the external magnetic field was applied in the in-plane direction parallel to the [100] axis or [110] axis of CMG. The insets show the enlarged hysteresis loops in the magnetization reorientation range of the magnetic field.

The saturation magnetization for the $L2_1$ ordered CMG was theoretically estimated to be about 4 Bohr magnetons per formula unit [23]. If we assume the density of the magnetic layer to be 8.4 g/cm$^3$, the theoretical saturation magnetization is about 780 emu/cm$^3$. However, the saturation magnetization was shown to decrease to about 600 emu/cm$^3$ when changing the ordering to B2 [24]. The values of saturation magnetization for all our samples were lower than the theoretical value for $L2_1$ phase (see Fig. 1). This is due to the fact that they contain mainly lower-order A2 and B2 phases.

The saturation magnetization was found to be about 625 emu/cm$^3$, 615 emu/cm$^3$ and 660 emu/cm$^3$, and the coercive field about 25 Oe, 20 Oe, and 30 Oe for $T_a$ = 300 °C, 400 °C, and 500 °C, respectively. Comparison of the in-plane hysteresis loops between the [100] and [110] crystallographic axes of CMG shows that the in-plane MCA is relatively weak, with an easy axis along the [110] direction for $T_a$ = 300 °C and 400 °C. In the case of $T_a$ = 500 °C, it was difficult to determine the direction of the easy axis from the magnetization curves. For accurate determination of the MCA, the FMR studies were performed.

For $T_a$ = 400 °C, and especially for $T_a$ = 500 °C, we observed characteristic constricted hysteresis loops. Different mechanisms can be responsible for such shapes of the hysteresis loops [25], and one of them is connected with sample inhomogeneity. In the case of our samples, the heterogeneity is most probably related to the coexistence of phases A2, B2 and $L2_1$. Micromagnetic simulations show that the reduced interface coupling can be responsible for such shapes of the hysteresis loops [26]. A simple relationship between coercive field and anisotropy can only be discussed in ideal crystals, and only in the case where domain structure does not occur. The presence of domain structure lowers the coercive field and also complicates the determination of anisotropy constants from magnetization curves. The situation is further complicated by the coexistence of different phases. On the other hand, ferromagnetic resonance is observed at much stronger fields (in our case, approximately 1000 Oe), where the domain structure disappears. Therefore, ferromagnetic resonance studies allow for a more accurate determination of magnetocrystalline anisotropy constants than comparing to magnetization curves for different crystallographic directions.

Fig. 2(a-c) shows the room temperature in-plane angular dependencies of the FMR resonance field ($H_{rez}$) for the samples annealed at $T_a$ = 300 °C (Fig. 2(a)), 400 °C (Fig. 2(b)), and 500 °C (Fig. 2(c)). For all the samples the [110] axis of CMG was parallel to the [100] (or equivalent) axis of the cubic MgO substrate. The in-plane variation of the resonance field is very small (about 20 Oe), compared to the average magnitude of $H_{rez}$ about 1000 Oe, which confirms the small in-plane MCA. Nevertheless, the angular dependence of the FMR resonance field reveals a four-fold symmetry characteristic (in this orientation) for the cubic epitaxial CMG layer. The easy axis of magnetization, corresponding to the minimum of $H_{rez}$, is parallel to the [110] axis of CMG for

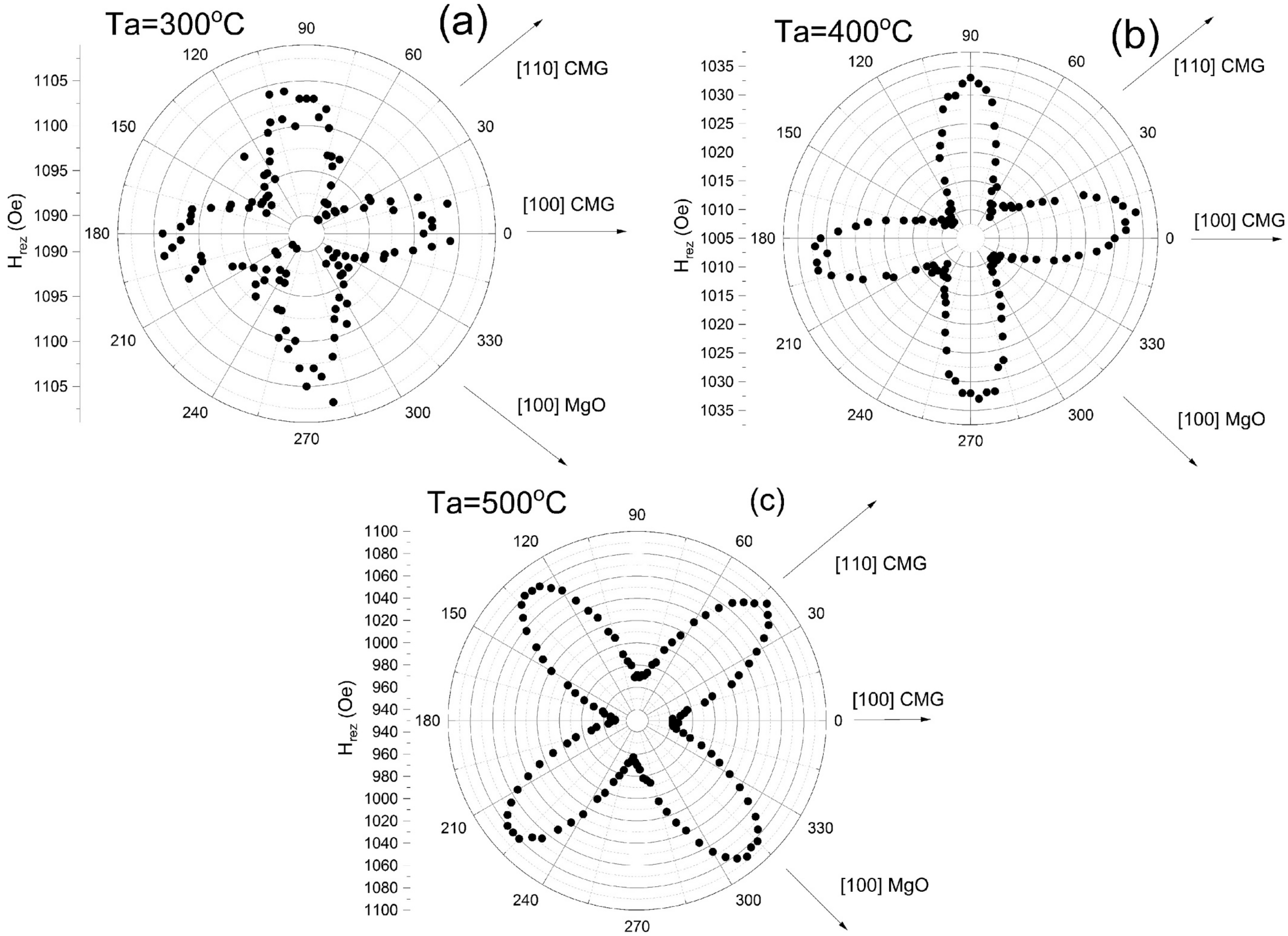

**Fig. 2.** The room temperature angular dependencies of the in-plane FMR resonance field (X-band 9.37 GHz) for the three samples. The samples were annealed at $T_a = 300$ °C (a), 400 °C (b) and 500 °C (c).

$T_a = 300$ °C and 400 °C. However, for $T_a = 500$ °C, it is changed and becomes parallel to the [100] axis of CMG.

Fig. 3 shows the room temperature angular dependencies of the out-of-plane FMR resonance field, for the three samples. The zero angle corresponds to the direction perpendicular to the film plane, and the 90-degree corresponds to the direction parallel to the [110] axis of CMG in the film plane. One can notice a slight increase in the resonance field in perpendicular to the film orientation with increasing annealing temperature, suggesting the changes of perpendicular MCA.

In our analysis, we used the same model as that presented earlier in Ref. [21], in which the free energy of the film was described by the equation:

$$F = -\sum_{i=1}^{3} M_i H_i + 2\pi M_s^2 \alpha_3'^2 + E_{mc} + E_{el} + E_{me}. \quad (1)$$

In Eq. 1, the first term represents the Zeeman energy, where $M_i$ and $H_i$ denote the components of magnetization and external magnetic field intensity, respectively. The second term accounts for the demagnetizing energy, with.

$M_s$ signifying the saturation magnetization and $\alpha_3'$ representing one of the direction cosines of magnetization associated with the axis perpendicular to the film. The $E_{mc}$, $E_{el}$, $E_{me}$ denote the magnetocrystalline, elastic and magnetoelastic energies, respectively.

The magnetocrystalline energy was assumed to be described by the formula:

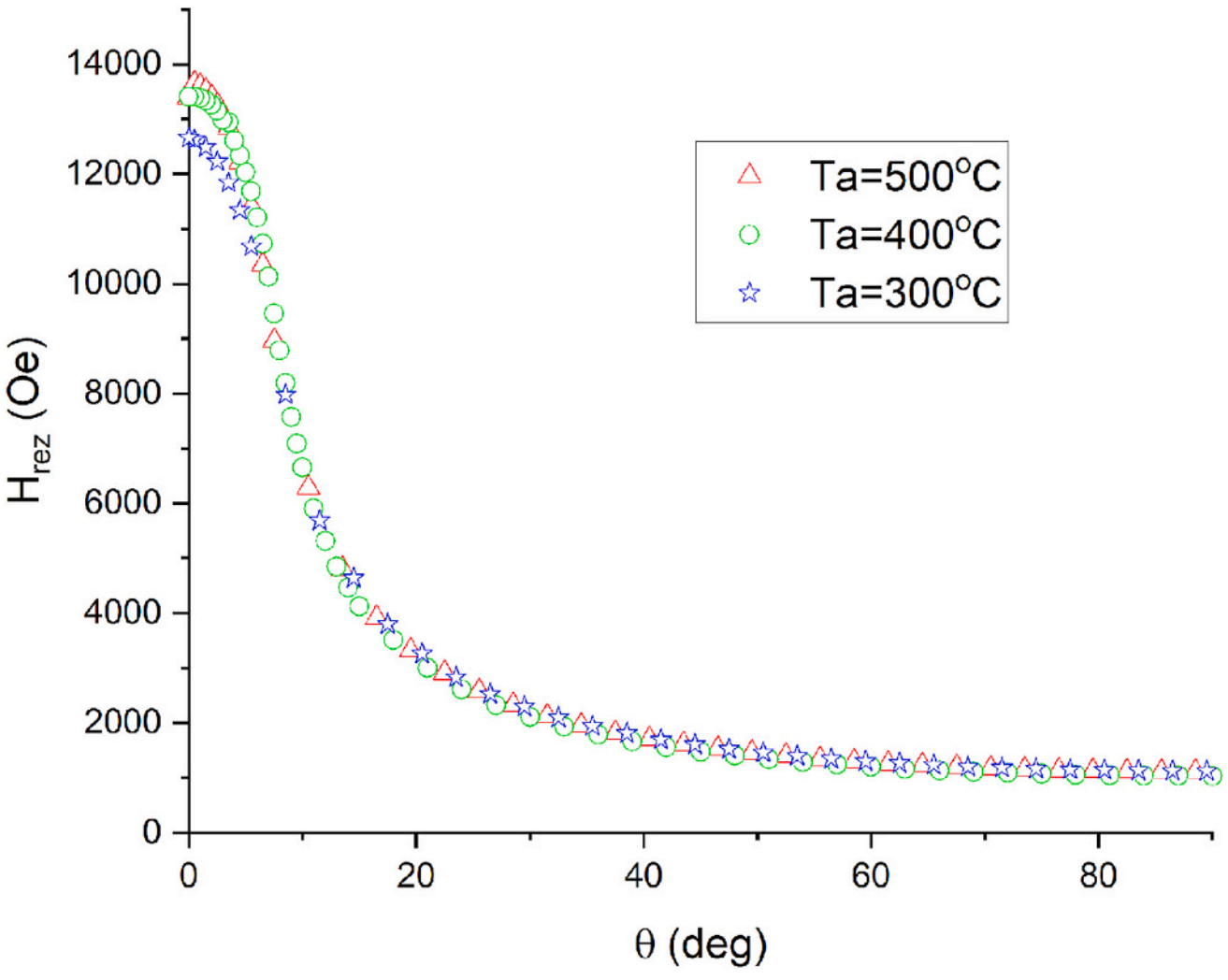

**Fig. 3.** The room temperature angular dependencies of the out-of-plane FMR resonance field (X-band 9.37 GHz) for the three samples. The samples were annealed at $T_a = 300$ °C (stars), 400 °C (circles), and 500 °C (triangles). The zero angle corresponds to the direction perpendicular to the film plane, and the 90-degree corresponds to the direction parallel to the [110] axis of CMG in the film plane.

$$E_{mc} = K_p\left(1-\alpha_3'^2\right) + K_1\left(\alpha_1'^2\alpha_2'^2 + \alpha_2'^2\alpha_3'^2 + \alpha_1'^2\alpha_3'^2\right), \tag{2}$$

where $\alpha_i$' are the direction cosines associated with the sample coordinate system, and the third coordinate is perpendicular to the film.

The $K_p$ and $K_1$ are the perpendicular and the first cubic MCA constants, respectively. The anisotropy constants were calculated using the magnitudes of the resonance field measured along the [100], [110], and [001] directions (see Fig. 2 and Fig. 3) and values of the magnitudes of saturation magnetization, using Eqs. 7–10 from Ref. [21]. The results of the calculations are given in Table 1.

The change of the sign of $K_1$ reflects the change of an easy axis of magnetization from [110] (negative value) to [100] (positive value). The negative sign of $K_p$ means that the perpendicular MC energy adds to the demagnetizing energy, making the in-plane orientation even more preferable.

The most significant change of $K_1$ was observed after increasing the annealing temperature up to 500 °C, which, according to the structural analysis presented in Ref. [13], can be associated with the appearance of the $L2_1$ phase (see Fig. 2 and Table 1 in Ref. [13]). For the samples annealed at 300 °C and 400 °C, $K_1$ is about $-0.4 \times 10^4$ erg/cm$^3$ and increases up to $2.2 \times 10^4$ erg/cm$^3$ for the sample annealed at 500 °C. The magnitude of $K_1$ of all samples is relatively low, i.e., of an order of $10^4$ erg/cm$^3$. The magnitude of $K_1$ for the CMG sample with highest content of the $L2_1$ phase ($2.2 \times 10^4$ erg/cm$^3$) is about twice smaller than for the similarly ordered $Co_2MnSi$ (about $-4 \times 10^4$ erg/cm$^3$) [21]. On the other hand for the similarly ordered $Co_2FeSi$ the magnitude of $K_1$ is very small (about $6 \times 10^3$ erg/cm$^3$), and when the content Fe/Ni in $Co_2Fe_xMn_{1-x}Si$ changes, the $K_1$ constant changes non-monotonic, reaching the largest magnitude (about $-8\ 10^4$ erg/cm$^3$) for the Fe content of about x = 0.4 [21].

In general, the magnitude of the room temperature cubic magnetocrystalline $K_1$ constant for Heusler alloys is relatively low (about order of $10^4$ erg/cm$^3$ [21]). It is usually the same order of magnitude as for nickel ($-4.5 \times 10^4$ erg/cm$^3$) and an order of magnitude lower than for iron ($4.8 \times 10^5$ erg/cm$^3$) [27]. Depending on composition, the $K_1$ in Heusler alloys can change its sign, and for certain compositions, very low values of $K_1$ can be obtained [21]. The present results show that the very low values of $K_1$ can also be due to the chemical ordering changes, and the highest magnitudes of $K_1$ are expected for the best ordered samples.

The perpendicular MCA constant for all investigated samples is negative by an order of $10^5$ erg/cm$^3$. The most remarkable change in $K_p$ was after increasing the annealing temperature up to 400 °C, i.e., with the appearance of the partially ordered B2 phase [13].

The results of the SMFMR studies are summarized in Table 2. $\Delta H_{100}$ and $\Delta H_{110}$ are the shifts of the FMR resonance lines, measured in-plane of the film in the direction [100] and [110], respectively. These shifts were induced by the strains $\varepsilon_{11}$ and $\varepsilon_{22}$ of the quartz rod to which the samples were glued. $\varepsilon_{11}$ and $\varepsilon_{22}$ denote the strains in the plane of the film in the direction perpendicular and parallel to the rod, respectively. As described in detail in Ref. [21], the strains were determined in each experiment using the calibration procedure to increase the accuracy of the magnetoelastic constants' determination.

The strains typically applied in our experiments ($\varepsilon_{11}$ about $-3 \times 10^{-6}$ and $\varepsilon_{22}$ about $1.5 \times 10^{-5}$) result in shifts of the resonance lines of about 0.5 Oe. The difference between $\Delta H_{100}$ and $\Delta H_{110}$ reveals the anisotropy of the magnetoelastic effect. One can see that for $T_a$ = 300 °C and 400 °C this anisotropy is relatively weak, and for $T_a$ = 500 °C, a very strong anisotropy of magnetoelastic properties was observed. The accuracy of ΔH and ε determination is about 5%. In our model (see Ref. [21] for details), we assumed the magnetoelastic energy of a cubic crystal to be described by two magnetoelastic constants $b_1$ and $b_2$ according to the formula [20]:

**Table 1**
The room temperature first cubic MCA constants, $K_1$, and the perpendicular MCA constant, $K_p$, for the samples annealed at the temperatures $T_a$.

| $T_a$ | $K_1$ ($10^4$ erg/cm$^3$) | $K_p$ ($10^5$ erg/cm$^3$) |
|---|---|---|
| 300 °C | $-0.4 \pm 0.05$ | $-4.5 \pm 0.2$ |
| 400 °C | $-0.4 \pm 0.05$ | $-7.4 \pm 0.2$ |
| 500 °C | $2.2 \pm 0.05$ | $-6.7 \pm 0.2$ |

**Table 2**
The shifts of the FMR lines measured using the SMFMR technique for the three samples, measured in the film plane with the external magnetic field applied parallel to the [100] axis ($\Delta H_{100}$) or [110] axis ($\Delta H_{110}$), caused by the periodic strains $\varepsilon_{11}$ and $\varepsilon_{22}$ of the polycrystalline quartz rod, to which the thin films were glued. Thus, both $\varepsilon_{11}$ and $\varepsilon_{22}$ are in-plane of the film [21].

| $T_a$ | $\Delta H_{100}$ (Oe) | $\varepsilon_{11}$ ($10^{-6}$) | $\varepsilon_{22}$ ($10^{-5}$) | $\Delta H_{110}$ (Oe) | $\varepsilon_{11}$ ($10^{-6}$) | $\varepsilon_{22}$ ($10^{-5}$) |
|---|---|---|---|---|---|---|
| 300 °C | 0.52 | −2.2 | 1.80 | 0.77 | −2.7 | 1.69 |
| 400 °C | 0.47 | −2.9 | 1.47 | 0.37 | −3.0 | 1.56 |
| 500 °C | 0.73 | −3.0 | 1.56 | −0.05 | −2.6 | 1.63 |

$$E_{me} = b_1\left(\alpha_1'^2\varepsilon_{11}' + \alpha_2'^2\varepsilon_{22}' + \alpha_3'^2\varepsilon_{33}'\right) + 2b_2\left(\alpha_1'\alpha_2'\varepsilon_{12}' + \alpha_2'\alpha_3'\varepsilon_{23}' + \alpha_1'\alpha_3'\varepsilon_{13}'\right), \tag{3}$$

where $\alpha_i$' and $\varepsilon_{ij}$' are direction cosines and strains in the coordinate system associated with the sample. The magnetoelastic constants, calculated in the framework of this model, are summarized in Table 3.

To calculate the $b_1$ and $b_2$ constants according to the model [21] the knowledge of the ratio of the elastic constants $c_{12}/c_{11}$ is necessary. Unfortunately, there is quite a large discrepancy between the elastic constants presented in the literature. One review of the data available can be found in the paper [28]. In our calculations, we used averaged values of all data presented in Ref. [28], i.e. $c_{11}$ = 281 GPa, $c_{12}$ = 157 GPa, $c_{44}$ = 132 GPa. The maximum deviations of the $c_{11}$, $c_{12,}$ and $c_{44}$ values given in Ref. [28] from our assumed mean values were 8%, 15% and 6%, respectively. The knowledge of the elastic constants is also necessary to calculate the magnetostriction constants. The longitudinal magnetostriction constants for the [100] and [111] directions, i.e. $\lambda_{100}$ and $\lambda_{111}$ can be calculated using the formulas (4) [21]:

$$\lambda_{100} = -\frac{2b_1}{3(c_{11}-c_{12})}, \qquad \lambda_{111} = -\frac{b_2}{3c_{44}} \tag{4}$$

For isotropic samples $b_1 = b_2$. Hence, the difference between magnetoelastic constants can be treated as a measure of the magnetoelastic properties' anisotropy. As can be expected based on the results of $\Delta H$ in Table 2 the largest difference between the $b_1 = -1.43\ 10^7$ erg/cm$^3$ and $b_2 = -2\ 10^5$ erg/cm$^3$constants is for the sample annealed at 500 °C, i.e. the sample containing 23% of $L2_1$ phase. For the highly $L2_1$ ordered sample, the two cubic magnetoelastic constants differ by almost two orders of magnitude, which indicates giant anisotropy of the magnetoelastic properties.

The calculated magnetostriction constants are also presented in Table 3. The enormous difference between the magnetoelastic constants $b_1$ and $b_2$ also causes an enormous difference between the calculated magnetostriction constants $\lambda_{100}$ and $\lambda_{111}$ [according to Eq. (4)], even increased due to the anisotropy of elastic properties. It is worth recalling here that for an isotropic sample $b_1 = b_2$, $c_{44} = 1/2\ (c_{11} - c_{12})$ and thus $\lambda_{100} = \lambda_{111}$ [21]. However, such calculated magnetostriction constants only refer to bulk materials. The actual magnetostriction, i.e. the changes of the sample dimensions in the external magnetic field, of the thin films is extremely small, because of the interaction of the thin magnetic film with a thick substrate.

**Table 3**
Room temperature magnetoelastic $b_1$ and $b_2$ and magnetostriction $\lambda_{100}$ and $\lambda_{111}$ constants of the CMG films annealed at Ta 300 °C, 400 °C, and 500 °C. The accuracy of the $b_1$ and $b_2$ determination is about 10%, and of $\lambda_{100}$ and $\lambda_{111}$ about 20%. Estimated higher error bar value for magnetostriction comes from a fact that elastic constants necessary for such calculations were taken from the literature but for bulk samples as the values for thin films are not available.

| $T_a$ | $b_1$ ($10^7$ erg/cm$^3$) | $b_2$ ($10^7$ erg/cm$^3$) | $\lambda_{100}$ ($10^{-6}$) | $\lambda_{111}$ ($10^{-6}$) |
|---|---|---|---|---|
| 300 °C | −0.92 | −1.37 | 4.9 | 3.5 |
| 400 °C | −0.91 | −0.71 | 4.9 | 1.8 |
| 500 °C | −1.43 | −0.02 | 7.7 | 0.05 |

The knowledge of the magnetoelastic constants and the magnetic layer distortion enables us to calculate the strain-induced magnetocrystalline anisotropy [22]. In the case of tetragonal distortion $\varepsilon_{11} = \varepsilon_{22}$ and $\varepsilon_{12} = \varepsilon_{23} = \varepsilon_{13} = 0$. Such distortion induces an additional axial anisotropy perpendicular to the film plane, described by the constant $K_{si} = b_1(\varepsilon_{11}\text{-}\varepsilon_{33})$, and $\varepsilon_{33} = -2(c_{12}/c_{11})\varepsilon_{11}$ [22]. In Ref. [15] the epitaxial CMG layers of the thicknesses changing from 10 nm to 80 nm deposited also directly on MgO (001) were investigated. The misfit strain was estimated to change from 3.62% for the 10 nm to 2.51% for the 80 nm layer, respectively. If we take $\varepsilon_{11} \approx 2.5\%$ and the average from our experiments $b_1 \approx -1 \times 10^7$ erg/cm$^3$, we obtain $K_{si} = -5.3 \times 10^5$ erg/cm$^3$, which is in quite good agreement with the $K_p$ values presented in Table 1. The magnitude of the first cubic magnetocrystalline anisotropy constant of our sample ($2.2 \times 10^4$ erg/cm$^3$) is also very close to the analogous value obtained in Ref. [15] for the 80 nm sample ($1.9 \times 10^4$ erg/cm$^3$). However, an easy axis of magnetization was in our case parallel to [100] direction, and in the case of the sample studied in Ref. [15] - parallel to [110] direction.

Let us now compare the magnetoelastic properties and magnetocrystalline anisotropy of the studied Weyl semimetal CMG thin films with those of the $Co_2Fe_xMn_{1-x}Si$ and $Co_2Fe_{0.4}Mn_{0.6}Si$ Heusler alloy thin films, which we investigated in our previous works [16,21,22,29,30]. The magnetoelastic constants ($b_1$, $b_2$) are all negative, corresponding to positive longitudinal magnetostriction constants. In most cases, these constants are of the order of $10^7$ erg/cm$^3$, which corresponds to magnetostriction constants of the order of $10^{-6}$. The only exception is the CMG layer containing (approx. 23%) the $L2_1$ phase, for which the magnetoelastic constant $b_2$ (and the magnetostriction constant $\lambda_{111}$) is two orders of magnitude smaller than for the other layers.

All samples are also characterized by relatively small cubic magnetocrystalline anisotropy constant $K_1$ (on the order of $10^4$ erg/cm$^3$) and relatively large negative perpendicular-to-film magnetocrystalline anisotropy constants $K_p$ (on the order of $10^5$–$10^6$ erg/cm$^3$) - see Table 1.

The most interesting seems to be the comparison of the ordered $Co_2MnGa$ and $Co_2MnSi$ samples. They differ only in one non-magnetic element (Ga/Si), but one of them is a Weyl semimetal and the other one is characterized by strong spin polarization. The differences in the magnetoelastic properties of these two materials are particularly noticeable. In the case of $Co_2MnGa$ sample containing about 23% of $L2_1$ phase, a great difference between the $b_1$ and $b_2$ magnetoelastic constants was observed (see Table 3 - sample annealed at 500 °C), while for the $Co_2MnSi$ sample containing similar amount of $L2_1$ phase the $b_1$ and $b_2$ constants have almost the same values (about $-1 \times 10^7$ erg/cm$^3$) [21]. One potential explanation for the observed difference may be correlated with topological properties of the $L2_1$ CMG phase. However, verifying this hypothesis requires additional research, because the anisotropy of the magnetoelastic properties of CMG does not necessarily have to be related to its topological properties. For example, in the case of the $Co_2Fe_xMn_{1-x}Si$ system (which is not Weyl semimetal) significant differences between the $b_1$ and $b_2$ constants can be achieved by substituting Mn ions with Fe ions [21]. With increasing Fe content increases the magnitude of $b_2$ increases (from about $-1 \times 10^7$ erg/cm$^3$ for $Co_2MnSi$ to about $-4 \times 10^7$ erg/cm$^3$ for $Co_2FeSi$), while the changes of the $b_1$ magnitude are rather moderate (it remains close to $-1$ $10^7$ erg/cm$^3$) [21]. It is also well known that for pure iron crystal the $b_1$ and $b_2$ constants, as well as $\lambda_{100}$ and $\lambda_{111}$ constants, have opposite signs ($\lambda_{100} = 2.0 \times 10^{-5}$ and $\lambda_{111} = -2.1 \times 10^{-5}$). The explanation of anisotropy of magnetoelastic properties for CMG requires calculations of changes in the band structure of the tested material under the influence of the strain. To the best of our knowledge, such calculations have not yet been performed.

The magnetoelastic energy described by the magnetoelastic constants [for cubic crystal $b_1$ and $b_2$ - see Eq. 3] tells us how the magnetocrystalline anisotropy changes with the strain. The anisotropy of magnetoelastic properties (in our case the difference between the constants $b_1$ and $b_2$) tells us that this dependence changes strongly with orientation of the strain tensor with respect to the crystallographic axes.

The results of our research suggest that band structure of the $L2_1$ ordered CMG alloy is very sensitive both to the crystallographic direction in which the sample is magnetized and to the orientation of the strain tensor with respect to the crystallographic axes. However, further research is required to verify whether this is directly related to the specific band structure of the Weyl semimetal.

Some works [16,17] indicate a connection between the magnetoelastic properties and intrinsic magnetic damping, which is important in the applications of these materials at microwave frequencies. It has been shown [17] that magnetic damping is enhanced when the magnetostriction constant is negative, whereas it is reduced when the constant is positive. This phenomenon originates from strain-induced modifications in the magnetization magnitude (via changes in exchange splitting), which, in turn, shift the density of states near the Fermi level, thereby affecting both magnetostriction and magnetic damping [17]. In the case of our samples, the magnetostriction constants are positive, which would suggest that the intrinsic magnetic damping associated with this effect should be rather moderate. A thorough investigation of this effect was difficult due to the significant inhomogeneous broadening of the resonance lines of the tested samples.

Our previous studies revealed the influence of magnetic layer thickness on both the magnetoelastic properties and the magnetocrystalline anisotropy constants. Similar effects are likely to be expected for the Weyl semimetal CMG. However, in the present work, all the investigated layers have the same thickness (75 nm). Therefore, we focused on the influence of chemical ordering on magnetocrystalline anisotropy and anisotropic magnetoelastic properties.

For thin films of certain Heusler alloys (such as CMG), achieving a high degree of chemical order may be challenging. In such cases, the degree and type of chemical ordering can significantly affect both the magnetocrystalline anisotropy and the magnetoelastic properties of these layers.

## 4. Conclusions

Changes in the chemical ordering in epitaxially grown $Co_2MnGa$ films upon annealing result in variations in both the magnetocrystalline anisotropy and magnetoelastic properties.

The most significant changes in both the cubic magnetocrystalline anisotropy and the magnetoelastic properties are associated with the presence of the highly ordered $L2_1$ phase (in our case 23% of the magnetic layer volume), which appears after annealing at 500 °C. The appearance of the $L2_1$ phase causes a significant increase in the magnitude of the magnetocrystalline cubic anisotropy constant (from $0.4 \times 10^4$ erg/cm$^3$ for A2 and B2 phases to $2.2 \times 10^4$ erg/cm$^3$), and the appearance of a giant difference between the two cubic magnetoelastic constants ($b_1 = -1.43 \times 10^7$ erg/cm$^3$, $b_2 = -2 \times 10^5$ erg/cm$^3$) or equivalent magnetostriction constants ($\lambda_{100} = 7.7 \times 10^{-6}$, $\lambda_{111} = 2 \times 10^{-8}$), which can be interpreted as a giant increase in the anisotropy of the magnetoelastic properties.

The magnetoelastic properties of the $Co_2MnGa$ films containing (having topological properties) $L2_1$ phase were compared with the properties of previously studied (non-topological) $Co_2Fe_xMn_{1-x}Si$ ($0 < x < 1$) films containing similar amount of $L2_1$ phase. Unlike in the case of $Co_2MnGa$, in the case of $Co_2MnSi$ the anisotropy of magnetoelastic properties is negligible ($b_1 \approx b_2 \approx -1 \times 10^7$ erg/cm$^3$). On the other hand, in the $Co_2Fe_xMn_{1-x}Si$ films the anisotropy of the magnetoelastic properties can be increased when substituting Mn by Fe ($b_1 \approx -1 \times 10^7$ erg/cm$^3$ and $b_2 \approx -4 \times 10^7$ erg/cm$^3$ for $Co_2FeSi$).

Our results suggest that the band structure of the $L2_1$ ordered

$Co_2MnGa$ alloy is very sensitive both to the crystallographic direction in which the sample is magnetized and to the orientation of the strain tensor with respect to the crystallographic axes. The possible connection of this fact with the specific band structure of the Weyl semimetal requires further studies.

The large out-of-plane axial MCA constant observed in our samples can be attributed to the combined effects of misfit strain and magnetoelasticity.

In Heusler alloy thin films, where achieving a high content of the ordered $L2_1$ phase is a challenge, the degree of chemical ordering may serve as a crucial factor in determining their magnetocrystalline anisotropy and magnetoelastic properties.

## CRediT authorship contribution statement

**O.M. Chumak:** Writing – review & editing, Investigation, Formal analysis, Conceptualization. **A. Nabiałek:** Writing – original draft, Project administration, Investigation, Formal analysis, Conceptualization. **T. Seki:** Writing – review & editing, Investigation, Conceptualization. **J. Wang:** Investigation. **K. Takanashi:** Writing – review & editing, Conceptualization. **L.T. Baczewski:** Writing – review & editing, Conceptualization. **H. Szymczak:** Writing – review & editing, Conceptualization.

## Declaration of competing interest

The authors declare that they have no known competing financial interests or personal relationships that could have appeared to influence the work reported in this paper.

## Acknowledgments

This work was supported by the National Science Centre of Poland, Project No. 2024/53/B/ST7/01848.

## Data availability

Data will be made available on request.